# Nonequilibrium Layer in the Crust of Neutron Stars and Nonequilibrium β-Processes in Astrophysics

## G. S. Bisnovatyi-Kogan

Space Research Institute, Russian Academy of Sciences, Moscow, and
MEPhI National Research Nuclear University, Moscow  and
Moscow Institute of Physics and Technology,  Dolgoprudnyi, Moscow region.

The formation of the chemical composition of neutron star envelopes, at densities $10^{10}$–$10^{13}$ g cm$^{-3}$, is considered. As hot matter is compressed in the process of collapse, which leads to the explosion of a core collapse supernova, the stage of nuclear equilibrium with free neutrino escape, kinetic equilibrium in β-processes, and, as a result, the establishment of limited nuclear equilibrium with a fixed number of nuclei takes place. Cold matter is compressed at a fixed number of nuclei whose atomic weight initially does not change and subsequently decreases. A pycnonuclear reaction of the fusion of available nuclei and a decrease in their number begin at the end. The compression of cold matter is accompanied by an increase in the mass fraction of free neutrons. In this case, the chemical composition of the envelope differs significantly from the equilibrium one and contains a considerable store of nuclear energy. Nonequilibrium β-reactions proceed at densities exceeding the upper bound for the non-equilibrium layer density, which lead to heating, nuclear energy release, and the possible attainment of a state of complete thermodynamic equilibrium. The thermodynamics of nonequilibrium β-processes, which lead to the heating of matter as neutrinos escape freely, is considered.

## 1. INTRODUCTION

Neutron stars were discovered in 1967 as sources of pulsed radio emission, pulsars [1]. This event occurred more than 30 years after the existence of such stars was predicted by the astronomers that investigated supernovae [2] and the theoreticians concerned with the interior structure of stars [3, 4]. A detailed history of the theoretical prediction of the possible existence of neutron stars is presented in [5]. After this outstanding discovery, the interest in neutron star physics increased sharply. The discovery of X-ray pulsars [6] confirmed the idea that X-ray sources could be associated with neutron stars and caused a new wave of activity in the investigation of neutron stars that continues until now. One of the main questions to which a large number of works were devoted is the explanation of the neutron star luminosity. The accretion model applied for neutron stars in binary systems [7] is the most developed one for explaining the X-ray luminosity of neutron stars. The radio emission from pulsars is explained by various processes in the magnetosphere of a rotating star [8]. Other works are devoted to the interior structure of neutron stars and determining the limit for their mass. In this paper we consider the problems related to the properties of matter in the neutron star crust at densities $10^{10}$ g cm$^{-3}$ ≤ ρ ≤ $10^{13}$ g cm$^{-3}$ and show that the physical phenomena important for various astrophysical applications take place there.

## 2. A BRIEF OVERVIEW OF THE NEUTRON STAR FORMATION PATH AT THE END OF NUCLEAR EVOLUTION

Neutron stars are currently believed to be formed in nonstationary processes related to supernova explosions. Explaining the observed energetics of supernovae was one of the main reasons for predicting the existence of neutron stars [2]. A study of the final stages of stellar evolution inevitably leads to the conclusion about the necessity of a hydrodynamic compression stage leading to the formation of a neutron star [9]. According to the theory of stellar evolution, the picture of the formation of neutron stars and black holes looks as follows. A degenerate carbon–oxygen core is formed in stars with masses $M \sim (8–12)$ $M_{solar}$ at the red giant stage, as a result of the reaction $3He^4 \rightarrow C^{12} + \alpha \rightarrow O^{16}$. The degenerate core growth process and the envelope mass loss proceed gradually until the core mass becomes close to the Chandrasekhar limit. For the molecular weight $\mu_e = A/Z = 2$, corresponding to the carbon–oxygen chemical composition, the Chandrasekhar limit is $M_{ch} \approx 5.8\ M_{solar} / \mu_e^2 \approx 1.45\ M_{solar}$. At a core mass greater than the limiting one an unrestrainable rise in the density and temperature occurs (the pressure of degenerate electrons is insufficient) and explosive carbon burning develops, whose result is a complete runaway with type Ia supernova explosion effects [10]. During the evolution of massive stars with $M > 12$ the same temperatures at which the carbon production and burning reactions proceed are reached at lower densities (the electrons are nondegenerate) and steady burning continues until the formation of a stellar core from iron-peak elements. Only nuclear reactions with the absorption of energy proceed during the subsequent evolution. The star loses its hydrodynamic stability in the process of energy loss through neutrino radiation and during the breakup of iron due to a high temperature. The decrease in adiabatic index $\gamma < 4/3$ due to the neutronization and breakup of iron peak nuclei is responsible for the loss of hydrodynamic stability [11]. Hot neutron stars are formed as a result of the collapse of such cores, which is accompanied by the envelope ejection and the appearance of a type II supernova. The discovery of neutron stars, pulsars, in supernova remnants observationally confirmed the conclusion that supernovae are associated with the birth of neutron stars. The discovery of the rapidly rotating pulsar PSR 0531+21 in a young supernova remnant called the Crab Nebula [12] was most important. The nebulae associated with supernova remnants were detected around the pulsars PSR 0833 – 45, 0611+22, 2021+51, 1154 – 62, etc. Thus, there is both theoretical and observational evidence for the formation of neutron stars as a result of the nonstationary process leading to a supernova explosion. Here we will consider the chemical composition of the envelopes of neutron stars as they cool down and the thermodynamics of matter in the presence of nonequilibrium β- processes accompanying the collapse and cooling of neutron stars.

A nonequilibrium layer, in which a store of energy up to $10^{49}$ erg is accumulated, is formed in the crust of a neutron star as it cools down [13, 14]. The nonequilibrium consists in a large excess of neutrons, which gives rise to superheavy nuclei near the stability boundary, $Q_n = 0$, and free neutrons. The slow evolution of the nonequilibrium layer through the diffusion of neutrons deep into the star [15] can be responsible for the long duration of the X-ray luminosity, $\sim 10^4$ yr. The nonstationary processes in a neutron star apparently caused by starquakes are accompanied by an abrupt decrease in the period [16, 17]. The subsequent relaxation, whereby the increase in the period returns to its mean value, is probably related to the interaction of superfluid vortices in the core of the neutron star with its crust [18]. A starquake causes the nonequilibrium matter to be brought outward, an explosive energy release, and could explain the observed γ-ray flares [19, 20] in and near the Galaxy. As the nonequilibrium matter flies apart, the interstellar medium is enriched by heavy elements.

3. PHYSICAL PROCESSES IN THE ENVELOPES OF NEUTRON STARS

One of the important evolutionary phases is the stage of a hot neutron star [21]. In collapse dynamics calculations [22], on reaching the nuclear densities corresponding to neutron stars, a temperature exceeding $2 \times 10^{11}$ K is reached in the matter. Various reactions that run very slowly at ordinary temperatures due to a high Coulomb barrier *B* can proceed in the matter under these conditions. This barrier is

$$B \approx \frac{Z_1 Z_2}{A^{1/3}} \text{ MeV}, \quad (1)$$

where *A* is the mean baryon number and *Z* is the nuclear charge. $B \approx 7$ MeV for the interaction of protons with iron-group elements. The mean thermal energy of the protons is

$$E_t = \frac{3}{2} kT = 0.133 T_9 \text{ MeV}, \quad T_9 = \frac{T}{10^9 \text{ K}}. \quad (2)$$

As a rule, the temperatures in stationary stars do not exceed $10^9$ K, $kT \ll B/10$, and, therefore, under stellar conditions the Coulomb barrier completely determines the rates of nuclear reactions with charged particles. If the temperatures are high enough for the rates of nuclear reactions to be great compared to the rates of other processes changing the density and temperature in the matter, then all nuclear reaction channels may be deemed open. In this case, the matter is under the conditions of detailed balance in nuclear reactions. For characteristic processes in stars the conditions of detailed balance in nuclear reactions are reached if the temperature exceeds $T > (3–5) \times 10^9$ K [11]. Nuclear reactions do not change the ratio of the total number of neutrons $N_n$ (in free and bound states) to the total number of protons $N_p$. Therefore, we can introduce the parameter $R_N = N_n/N_p$ and consider the nuclear equilibrium at a given $R_N$ [23]. However, in real physical situations $R_N$ changes as a result of β-processes and, in general, depends on time. Under some conditions it is possible to eliminate the dependence on time, for example, in the conditions of complete thermodynamic equilibrium, when there is detailed balance in β-processes as well [24]. A situation where the conditions of complete thermodynamic equilibrium are difficult to fulfil is encountered most often in nature. This is because of the huge difference in the time scales of the processes in nuclear and weak interactions, $\tau_n \ll \tau_\beta$ at $T > 5 \times 10^9$, and the absence of equilibrium for a particular species of particles (neutrinos) involved in the processes. Under these conditions, by separating the processes into fast and slow ones, we can consider the detailed balance in fast processes and the kinetic equations for slow processes. An example of such a situation arises at the collapse stage of a star at densities $\rho < 10^{12}$ g cm$^{-3}$ and temperatures $5 \times 10^9$ K $< T < 10^{10}$ K. The matter is transparent to neutrinos and there is no equilibrium for it, while there exists detailed balance for nuclear reactions. An important special case is the kinetic equilibrium in β-processes [25, 26], when the total rate of decrease in $R_N$ is equal to the total rate of its increase, i.e., $R_N$ remains constant.

## 4. THE FORMATION OF A NONEQUILIBRIUM LAYER IN THE NEUTRON STAR ENVELOPE

### 4.1. Cooling of the Neutron Star Envelope after the Supernova Explosion

When the temperature in the envelope becomes lower than $T \leq 4 \times 10^9$ K, the time of the reactions with charged particles begins to exceed the cooling time. At densities $\rho \geq \sim 10^9$ g cm$^{-3}$ the matter is composed of nuclei, neutrons, and relativistically degenerate electrons. Reactions with neutrons, neutron photodetachment and capture, β-decays at $\varepsilon_\beta \geq \varepsilon_{f,e}$, and $e^-$ captures at $\varepsilon_\beta \leq \varepsilon_{f,e}$ can proceed under these conditions. Here, $\varepsilon_\beta$ is the nuclear β-decay energy and $\varepsilon_{f,e}$ is the electron Fermi energy defined as follows:

$$\varepsilon_{f,e} = m_e c^2 \left[ \frac{\rho}{\mu_e \times 10^6 \text{ g/cm}^3} \right]^{1/3} \text{ erg,} \tag{3}$$

where $\mu_e = 1/\sum_i (Z_i/A_i) x_i$ is the number of nucleons per electron and $x_i$ is the mass fraction of element $i$. We will estimate the characteristic photodetachment time of a neutron $\tau_{\gamma n}$ as a function of its binding energy in the nucleus from the formula [27]

$$\frac{1}{\tau_{\gamma n}} = 6.95 \times 10^{33} T_9^{3/2} \langle \overline{\sigma}_{th} \rangle \times \exp\left(-\frac{11.605 Q_{n6}}{T_9}\right) \text{ s}^{-1}. \tag{4}$$

Here, $Q_{n6}$ MeV is the binding energy of the neutron in the nucleus and $\langle \overline{\sigma}_{th} \rangle$ is the cross section for the $A(n, \gamma)(A + 1)$ reaction averaged over the Planck distribution of photons. For $\langle \overline{\sigma}_{th} \rangle \sim 1$ b and $T_9 \leq 0.04 Q_{n6}$ we obtain $\tau_{\gamma n} > 10^7$ yr, i.e., at $T_9 < 0.4$ the neutrons with a binding energy $Q_n > Q_{nb} = 10$ MeV are virtually undetachable from the nucleus. Thus, there is equilibrium in reactions with neutrons in the system for nuclear with a binding energy of the last neutron below 10MeV.

The formation of a nonequilibrium chemical composition is seen on the diagram in Fig. 1. The nuclear fission and stability boundaries are plotted on the $(A, Z)$ plane. Let us trace the track of change in $(A, Z)$ for a nucleus surrounded by neutrons that constitute half the mass fraction of the matter or more. The $(A, Z)$ plane is broken down into three regions:

region I: $Q_n > Q_{nb}$,

region II: $0 < Q_n < Q_{nb}$ and $\varepsilon_\beta > \varepsilon_{f, e}$,

region III: $0 < Q_n < Q_{nb}$ and $\varepsilon_\beta < \varepsilon_{f, e}$.

At a high concentration of neutrons the nuclei with a large $Q_n$ will capture neutrons and will pass from region I into regions II and III irrespective of $\varepsilon_\beta$. Nuclei with a large excess of neutrons located far from the nuclear stability "valley" are formed in this case. In region III there is equilibrium in reactions with neutrons; the β-processes of the capture of electrons with a high Fermi energy always lead to a decrease in $Z$ due to the inequality $\varepsilon_\beta < \varepsilon_{f, e}$ and, as a result, all of the nuclei from this region will reduce $Z$ and will fall into the region below the level $ab$. On the other hand, in region II the equilibrium in reactions with neutrons is accompanied by $e^-$ β-decays with increasing $Z$ until all of the nuclei fall into the region above the level $cd$.

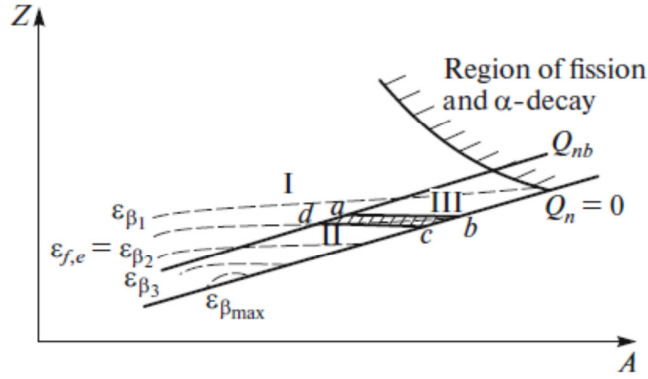

**Fig. 1.** The formation of chemical composition at the stage of limited equilibrium. The $Q_n = 0$ line separates the region of existence of nuclei; the $Q_{nb}$ line separates region I, where no neutron photodetachment is possible, from regions II and III. The dashed lines indicate the levels of constant $\varepsilon_\beta = Q_p - Q_n$, $\varepsilon_{\beta 1} < \varepsilon_{\beta 2} < \ldots < \varepsilon_{\beta max}$. $Q_n > Q_{nb}$ in region I; $Q_n < Q_{nb}$ and $\varepsilon_{f,e} > \varepsilon_\beta$ in region II. $Q_n < Q_{nb}$ and $\varepsilon_{f,e} > \varepsilon_\beta$ in region III. The hatched region *abcd* specifies the boundaries of $(A, Z)$ in the case of limited equilibrium with given $Q_{nb}(T)$ and $\varepsilon_{f,e}(\rho)$. The line with hatching on the right separates the region of fission and α-decay from [14].

We see that under the conditions of limited equilibrium the chemical composition of the matter is determined by the rather narrow region in $A$ and $Z$, bounded by the boundaries *abcd*. There is no exit from this region due to the absence of admissible β-processes and neutron photodetachment. At $Q_{nb} < 1$ MeV and $T_9 < 0.4$ there remains only one nucleus located at the stability boundary, for which

$$\varepsilon_\beta = Q_p - Q_n \approx Q_p = \varepsilon_{f,e}. \tag{5}$$

Additional processes during the formation of neutron excess nuclei are their fission, α-decay, and pycnonuclear reactions like $2(A, Z) = (2A, 2Z)$ [28]. To estimate the influence of the α-decay and fission of nuclei, let us consider the properties of the nuclei far from the stability valley. So far there are no rigorous theory and calculation of the parameters of such nuclei and, a fortiori, experimental data on the heavy nuclei deep in the instability region. Interpolating the estimates by P.E. Nemirovsky, we will roughly assume that the boundary of nuclear stability with respect to neutron evaporation passes at $A = 4Z$ for $Z > 6$ and the proton binding energy at this boundary is [14, 29, 30]

$$Q_{pb} = \left(33 - \frac{Z}{7}\right) \text{ MeV}. \tag{6}$$

At $Z < 6$, the value of $Q_{pb}$ rapidly decreases. For cold matter, given (5) and (6), we obtain an expression for the nuclear charge $Z_0$ at the neutron stability boundary as

$$Z_0 = 7(33 - \varepsilon_{f,e}) = 7\left[33 - 0.511\left(\frac{\rho_6}{\mu_e}\right)^{1/3}\right],$$

$$A_0 = 4Z_0. \tag{7}$$

Here, $\varepsilon_{f,e}$ is given in MeV and $\rho_6 = \rho/10^6$ g cm$^{-3}$. Suppose that the concentration of neutrons is so high that there are still many free neutrons in the matter even when all nuclei are at the stability boundary. At $\varepsilon_{f,e} > 33$ MeV in our approximation there are no stable nuclei and the capture of electrons is a nonequilibrium one; as a result, rapid heating occurs [31–33], new seed nuclei are formed, and the matter comes to a state close to nuclear equilibrium [14]. The inner boundary of the nonequilibrium layer lies approximately at density $\rho_2$ and pressure $P_2$:

$$\rho_2 \approx \mu_e \times 10^6 \left(\frac{33}{0.511}\right)^3 \approx 2.7 \times 10^{11} \mu_e \text{ g cm}^{-3}, \tag{8}$$

$$P_2 \approx 1.2 \times 10^{15} \left(\frac{\rho_2}{\mu_e}\right)^{4/3} \approx 2 \times 10^{30} \mu_e \text{ dyne cm}^{-2}. \tag{9}$$

In this case, $\mu_e \approx 4$ and $\rho_2 \approx 10^{12}$ g cm–3. The outer boundary of the nonequilibrium layer is defined by the boundary of stability for nuclei with $Q_n = 0$ with respect to fission. Using an empirical formula for the half-life, we find [14] that at $Z < 153 = Z_{max}$ the half-life exceeds $3 \times 10^7$ yr. Taking $Z_1 = Z_{max}$ at the outer boundary of the layer, we obtain the parameters

$$\rho_1 \approx \mu_e \times 10^6 \left(\frac{11}{0.511}\right)^3 \approx 10^{10} \mu_e \text{ g cm}^{-3}, \tag{10}$$

$$P_1 \approx 1.2 \times 10^{15} \left(\frac{\rho_2}{\mu_e}\right)^{4/3} \approx 2.5 \times 10^{28} \text{ dyne cm}^{-2}. \tag{11}$$

At $\mu_e \approx 4$ we have $\rho_1 \approx 4 \times 10^{10}$ g cm$^{-3}$. The layer mass $M_l$ is estimated from the relation that follows from the thin-layer equilibrium condition, for neutron stars with masses of (1.4–0.5) .

$$M_l = \frac{4\pi r^4}{GM} P_2 \approx (0.1-1) P_2 \approx 2 \times (10^{29} - 10^{30}) \text{ g}$$
$$= (10^{-4} - 10^{-3}) M_\odot, \tag{12}$$

Lower mass neutron stars have a larger radius and a thicker and more massive nonequilibrium envelope. Let us assume the energy $Q$ being released when a mixture of neutrons and superheavy nuclei passes into an equilibrium state to be $Q \approx 3 \times 10^{-3} M c^2$. The total store of energy in the nonequilibrium layer will then be $Q_{tot} = \eta M_l c^2 \approx (10^{48} - 10^{49})$ erg for neutron stars with masses of (1.4–0.5) , respectively. Thus, a neutron star possesses a store of energy that makes the manifestation of its observed activity possible without external energy sources.

*4.2. Compression of Cold Matter during Accretion*

The chemical nuclear composition of cold mater corresponding to the minimum of energy at a given density was calculated in [34]. At a density $\rho \leq 3 \times 10^9$ g cm$^{-3}$ the Fe$^{56}$ nucleus corresponds to the minimum energy. As the density rises, the electron Fermi energy $\varepsilon_{f,\,e}$ increases and two electron capture reactions occur successively on reaching $\varepsilon_{f,\,e} = Q_p - Q_n = \varepsilon_{\beta AZ}$, because the binding energy of even–even nuclei is greater than that of even–odd ones:

$$(A,Z) + e^- \rightarrow (A, Z-1) + \nu,$$
$$(A, Z-1) + e^- \rightarrow (A, Z-2) + \nu, \tag{13}$$

$$Fe^{56} + e^- \to Mn^{56} + \nu,$$
$$Mn^{56} + e^- \to Cr^{56} + \nu. \quad (14)$$

The second reaction of the chain is a nonequilibrium one and is accompanied by heating (see Section 5). If the heating is not enough to overcome the Coulomb barrier, then as the density rises, the neutronization process at constant $A$ will continue until the appearance of a nucleus at the neutron stability boundary with $Q_n = 0$. As the density rises further, the capture of an electron by a nucleus is accompanied by the emission of one or more neutrons:

$$(A, Z) + e^- \to (A - k, Z - 1) + kn + \nu. \quad (15)$$

This process continues along the curve on the $(A, Z)$ plane with $Q_n = 0$ until a nucleus with the maximum available $Q_p$ appears. According to the estimates from [30], $Q_{p\max} \approx 33$ MeV for the carbon nucleus with $Z = 6$ and $A = 22$. As was pointed out in [14], the formation of a stable nucleus is impossible as the density rises further and, therefore, the nuclear composition will be restructured and the transition to a new stable state will occur during further neutronization. Since the β-processes are highly nonequilibrium ones, intense heating will occur and the transition to a state close to complete nuclear equilibrium is possible. As was found in [31], for a highly nonequilibrium β-reaction with $\varepsilon_{f,e} \gg \varepsilon_{\beta AZ}$ about $(1/6)\varepsilon_{f,e}$ turns into heat. At $\varepsilon_{f,e} = 33$ MeV approximately 6 MeV per reaction turns into heat; this corresponds to $T \approx 5 \times 10^{10}$ K at which the Coulomb barrier can be overcome and the transition to an equilibrium composition is possible. The carbon composition with $Z = 6$ and $A = 22$ corresponds to a density $\rho_2 \sim 10^{12}$ from (8). The evolution of the chemical composition of cold matter under compression, given the action of nuclear forces and other factors in the mixture of nuclei and neutrons, was accurately calculated in [28]. In comparison with [14], the pycnonuclear nuclear fusion reactions that begin at a density of $1.4 \times 10^{12}$ g cm$^{-3}$ for the composition $A = 32$ and $Z = 9$ were taken into account in [28]. A nucleus with $A = 64 - k$, $Z = 18$ and $k$ free neutrons appear as a result of the pycnonuclear reaction. As a result of the further compression, the capture of electrons occurs with the evaporation of neutrons up until the formation of a nucleus with $A = 33$ and $Z = 9$ at a density of $3.3 \times 10^{12}$ g cm$^{-3}$. The calculation of such a stepwise process then continued up to a density of $5 \times 10^{13}$ g cm$^{-3}$. It is also noted in [28] that the process is accompanied by nonequilibrium heating, which can lead to a state of complete equilibrium, but this was not considered in the calculations. Various aspects of the structure and physical properties of matter in the crust of neutron stars are outlined in the reviews [35, 36].

5. HEATING IN NONEQUILIBRIUM WEAK INTERACTION REACTIONS

If the electron Fermi energy reaches the energy difference $\Delta_{12}$ at a low temperature, then the β-capture of an electron (neutronization) occurs with the transition between the ground states of nuclei $(A, Z)$ and $(A, Z - 1)$. For example, for the Fe$^{56}$ nucleus neutronization with the formation of an Mn$^{56}$ nucleus begins at $\varepsilon_{f,e} = \Delta_{12} \approx 3.8$ MeV. In this case, the process is an equilibrium one and all the energy of the captured electron is converted into the rest energy of the nucleus. During rapid stellar collapse the electrons have no time to be captured by nuclei and a situation where $\varepsilon_{f,e} \gg \Delta_{12}$ arises. In this case, the capture of electrons is a nonequilibrium one, the entropy increases, and the matter heats up. The rise in temperature under nonequilibrium neutronization is illustrated in Fig. 2, where it is shown how the step in the distribution of electrons with $T = 0$ is smeared and the electrons acquire a nonzero temperature. The nonequilibrium heating during β-captures was first investigated in [31], was independently considered shortly afterwards in [33], and subsequently in [32, 37].

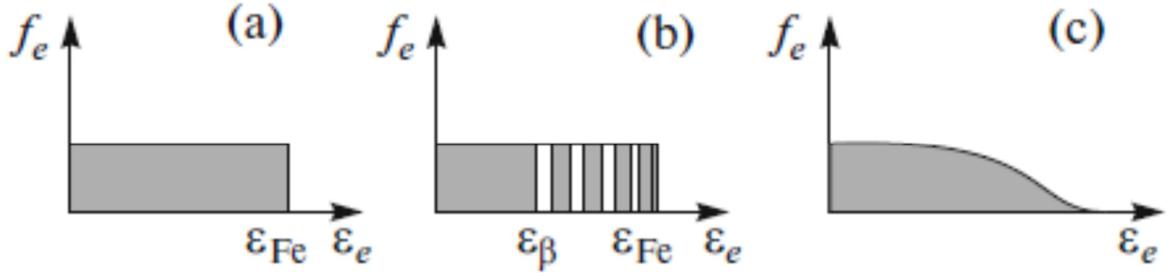

**Fig. 2.** Heating of cold matter during a nonequilibrium β-capture: (a) the electron distribution function before neutronization, $T = 0$; (b) the electron distribution function after nonequilibrium captures; (c) the electron distribution function after relaxation, $T > 0$.

The expression for the heating rate under nonequilibrium neutronization follows from the second law of thermodynamics for a variable composition of particles [38]:

$$dE - \frac{P}{\rho^2}d\rho = TdS + \sum_i \mu_{ti} dn_i = dQ = -\frac{Q_\nu}{\rho} dt. \tag{16}$$

Here, $S$ is the total entropy, $\mu_{ti}$ is the thermodynamic potential for particles of species $i$, and $\delta = \Delta_{12}/m_e c^2$. For the capture of cold electrons we obtain the reaction rate $W_e$ and the neutrino loss rate $Q_\nu$ [39]:

$$W_e = \frac{g_{Z'}}{g_Z} \frac{\ln 2}{(Ft_{1/2})_{Z'}} \int_\delta^{u_{fe}} (u^2 - 1)^{1/2} u(u-\delta)^2 du$$

$$= \frac{g_{Z'}}{g_Z} \frac{\ln 2}{(Ft_{1/2})_{Z'}} [F_0(u_{fe}) - F_0(\delta)] \tag{17}$$

reactions per nucleus in a second,

$$Q_\nu = \frac{g_{Z'}}{g_Z} \frac{\ln 2}{(Ft_{1/2})_{Z'}} m_e c^2 \int_\delta^{u_{fe}} (u^2 - 1)^{1/2} u(u-\delta)^3 du$$

$$= \frac{g_{Z'}}{g_Z} \frac{\ln 2}{(Ft_{1/2})_{Z'}} m_e c^2$$

$$\times [F_\epsilon(u_{fe}) - F_\epsilon(\delta)] \text{ erg s}^{-1} \text{ per nucleus.} \tag{18}$$

Here, $g_Z$ and $g_{Z'}$ are the statistical weights of the initial and final nucleus, respectively, $Z' = Z - 1$, and $u_{fe} = (\varepsilon_{f,e}/m_e c^2) + 1$. Analytical expressions for the Fermi function

$$F_0(x) = \int_1^x (u^2 - 1)^{1/2} u(u-\delta)^2 du \tag{19}$$

and the thermal Fermi function

$$F_\epsilon(x) = \int_1^x (u^2 - 1)^{1/2} u(u - \delta)^3 du \tag{20}$$

are given in [39]. For nonrelativistic nuclei and cold electrons we have

$$\mu_{t,Z} \approx m_{A,Z} c^2, \quad \mu_{t,Z-1} \approx m_{A,Z-1} c^2,$$
$$\mu_{te} \approx \varepsilon_{f,e} + m_e c^2. \tag{21}$$

From (16), given (17)–(21), we obtain
(

$$\rho T \frac{dS}{dt} = [(\varepsilon_{f,e} + m_e c^2 - \Delta_{Z'Z}) W^{(a)} - Q^{(a)}] n_{A,Z}$$
$$= \frac{g_{Z'}}{g_Z} \frac{\ln 2}{(Ft_{1/2})_{Z'}} n_{A,Z} m_e c^2$$
$$\times \{(u_{fe} - \delta)[F_0(u_{fe}) - F_0(\delta)] - F_\epsilon(u_{fe}) + F_\epsilon(\delta)\}. \tag{22}$$

For $u_{fe} \gg \delta \gg 1$ we have

$$F_0(u_{fe}) \approx \frac{u_{fe}^5}{5}, \quad F_0(\delta) \approx \frac{\delta^5}{5},$$
$$F_\epsilon(u_{fe}) \approx \frac{u_{fe}^6}{6}, \quad F_\epsilon(\delta) \approx \frac{\delta^6}{6},$$
$$\rho T \frac{dS}{dt} \approx \frac{g_{Z'}}{g_Z} \frac{\ln 2}{(Ft_{1/2})_{Z'}} n_{A,Z} m_e c^2 \frac{u_{fe}^6 - \delta^6}{30}. \tag{23}$$

We see that 1/6 of the energy goes into heating, on average, under highly nonequilibrium conditions, while the neutrino radiation carries away 5/6 of the energy of the captured electrons. Nonequilibrium neutronization at a finite temperature also leads to an increase in entropy, but this effect decreases with rising temperature at fixed $\varepsilon_{f,e}$. As the temperature rises further, heating is replaced by cooling, among other things, due to the opening of additional neutrino production and free runaway channels. The neutronization thermodynamics at a finite matter temperature was investigated in [40].

At $u_{fe} \gg \delta \gg 1$, when the energy of electrons is much greater than the mass difference of nuclei in the ground state, the electron capture to excited state of the final nucleus becomes energetically possible. Heat is released when the excitation is removed, which increases the thermal effect of nonequilibrium neutronization. In [41] the heating was calculated by taking into account the transitions to an excited state in the simplest model of nuclei composed of a Fermi gas of neutrons and protons. The mean energy of the outgoing neutrinos was found to be $\varepsilon_\nu \approx 0.6\varepsilon_{f,e}$ versus $(5/6)\varepsilon_{f,e}$ during the transitions between the ground states. Accordingly, approximately $0.4\varepsilon_{f,e}$ goes into heating and the increase in entropy, on average, compared to $(1/6)\varepsilon_{f,e}$ in the case of two ground states of the nucleus. The capture to an excited state of the final nucleus is illustrated in Fig. 3c.

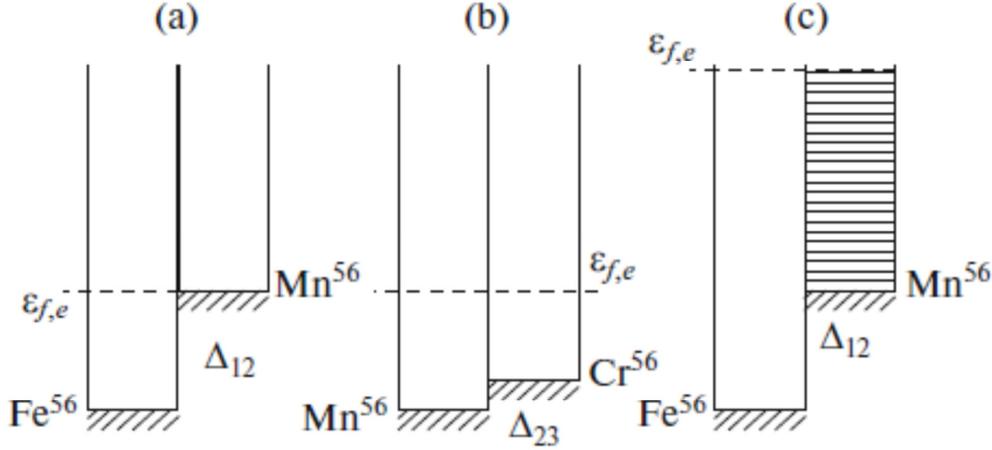

**Fig. 3.** Heating of cold matter during a nonequilibrium β-capture: (a) an equilibrium electron β-capture at $\varepsilon_{f,e} \approx \Delta_{12}$; (b) a nonequilibrium transition between the ground states at $\varepsilon_{f,e} \gg \Delta_{12}$; (c) a nonequilibrium transition between the ground state of the initial nucleus and all of the admissible states of the final nucleus, when the excitation energy $\varepsilon_{ex} \leq \varepsilon_{f,e} - \Delta_{12}$.

## 6. ASTROPHYSICAL MANIFESTATIONS

The nonequilibrium heating of a neutron star during accretion in a binary system has been considered in many papers (see, e.g., [42]). The formation of a nonequilibrium layer during accretion onto a neutron star can lead to deviations from axial symmetry in the neutron star and gravitational radiation. It was hypothesized in [43] that this gravitational radiation limits the increase in the period during disk accretion, which explains the observed limitation of the angular velocity of millisecond pulsars.

The nonequilibrium heating during the collapse of a star with the formation of a neutron star was first taken into account in the calculations of the compression of initially cold matter in [44, 45]. After the appearance of [37], the heating in nonequilibrium β- processes was taken into account in almost all of the theoretical studies of core-collapse supernovae (see, e.g., [46, 47]).

The maintenance of the luminosity of a massive white dwarf during neutronization and the formation of a tiny new-phase chromium core with a relative mass $M_{new}/M \leq 3 \times 10^{-3}$, whereby the stability with respect to collapse is retained, was consider in the first paper on heating as a result of nonequilibrium neutronization [31]. As the white dwarf cooling calculations showed, given the nonequilibrium heating, the cooling time at the final stages can increase severalfold [48]. This is true only for massive white dwarfs in the above range of masses. The neutronization of an iron nucleus occurs in two steps, the second of which is a nonequilibrium one (see Figs. 3a and 3b).

During the formation of a nonequilibrium layer in a young neutron star, apart from superheavy nuclei, neutrons are produced in it. Under the action of gravity free neutron diffuse into the star through its crust, in which a Coulomb crystal is formed. The release of heat during the diffusion of neutrons slows significantly the cooling of the neutron star, maintaining its luminosity at a level of $10^{36}$–$10^{34}$ erg s$^{-1}$ for ~$10^4$ yr [15].

The existence of a nonequilibrium layer in a young neutron star can play an important role in explaining the nonstationary behavior of objects known as soft gamma repeaters. The irregular flares observed from them are often explained by the release of energy during the annihilation of an anomalously strong magnetic field $B \sim 10^{15}$ G in neutron stars called magnetars. An alternative model is related to the flares resulting from a nuclear explosion during the fission of superheavy elements brought out to the surface as a result of the development of instabilities leading to a starquake. A schematic view of the explosion is presented in Fig. 4 (see [49, 50]). This model was considered to explain the cosmic gamma-ray bursts in 1975 [19] long before their cosmological origin was discovered.

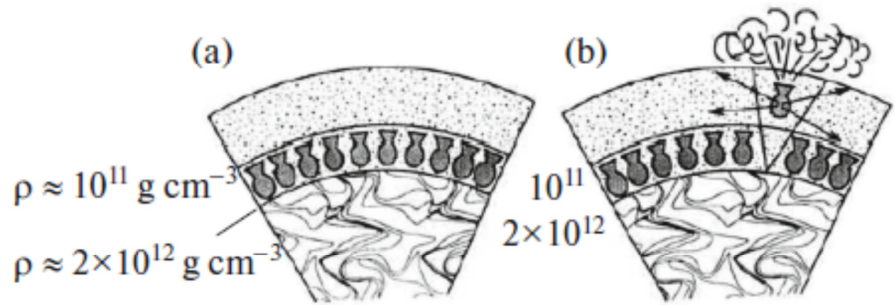

**Fig. 4.** Schematic view of the nonequilibrium layer in a neutron star: (a) in quiescence and (b) after a starquake and a nuclear explosion.

FUNDING

This work was supported by the Russian Science Foundation (grant no. 18-12-00378).